\pdfoutput=1

\documentclass[journal]{IEEEtran}
\usepackage{cite}
\ifCLASSINFOpdf
\else
\fi
\usepackage{graphicx}
\usepackage{caption}
\usepackage[justification=centering]{caption}

\usepackage{subfigure}

\usepackage{amsmath}
\usepackage{algorithm}
\usepackage{algorithmic}

\usepackage{array}

\usepackage{stfloats}
\usepackage{url}
\usepackage{caption}

\begin{document}
\title{Automated Classification of Seizures against Nonseizures: A Deep Learning Approach}

\author{Xinghua Yao,
        Qiang Cheng$^{*}$,
        and Guo-Qiang Zhang$^{*}$
\thanks{* Corresponding authors.}
\thanks{Xinghua Yao and Qiang Cheng are with the Institute of Biomedical Informatics, University of Kentucky, Lexington, Kentucky 40536-0082, USA (e-mail: xhyaosues@aliyun.com; Qiang.Cheng@uky.edu).}
\thanks{Guo-Qiang Zhang is with The University of Texas Health Science Center at Houston, Houston, Texas 77030, USA (e-mail: gqatcase@gmail.com).}
}


\markboth{IEEE Transactions on Biomedical Engineering}%
{Shell \MakeLowercase{\textit{et al.}}: Bare Demo of IEEEtran.cls for IEEE Journals}




\maketitle

\begin{abstract}
In current clinical practice, electroencephalograms (EEG) are reviewed and analyzed by well-trained neurologists to provide supports for therapeutic decisions. The way of manual reviewing is labor-intensive and error prone. Automatic and accurate seizure/nonseizure classification methods are needed. One major problem is that the EEG signals for seizure state and nonseizure state exhibit considerable variations. In order to capture essential seizure features, this paper integrates an emerging deep learning model, the independently recurrent neural network (IndRNN), with a dense structure and an attention mechanism to exploit temporal and spatial discriminating features and overcome seizure variabilities. The dense structure is to ensure maximum information flow between layers. The attention mechanism is to capture spatial features. Evaluations are performed in cross-validation experiments over the noisy CHB-MIT data set. The obtained average sensitivity, specificity and precision of 88.80\%, 88.60\% and 88.69\% are better than using the current state-of-the-art methods.
In addition, we explore how the segment length affects the classification performance. Thirteen different segment lengths are assessed, showing that the classification performance varies over the segment lengths, and the maximal fluctuating margin is more than 4\%. Thus, the segment length is an important factor influencing the classification performance.
\end{abstract}

%Extensive experiments over the noisy data of CHB-MIT are conducted to evaluate our proposed approach.
%In cross-validation experiments, the obtained result of (Sensitivity, Specificity, F1 Score, Precision, Accuracy) is (87.00\%$\pm$0.0363, 88.60\%$\pm$0.0463, 87.71\%$\pm$0.0228, 88.63\%$\pm$0.0388, 87.80\%$\pm$0.0230). The achieved result in cross-patient experiments is (83.72\%$\pm$0.1349, 84.06\%$\pm$0.1379, 83.63\%$\pm$0.0888, 85.36\%$\pm$0.1020, 83.89\%$\pm$0.0833).

\begin{IEEEkeywords}
Independently RNN, dense structure, attention mechanism, seizure/nonseizure classification, electroencephalogram.
\end{IEEEkeywords}

% \fi
\IEEEpeerreviewmaketitle

\section{Introduction}

\IEEEPARstart{E}{pilepsy} is a chronic neurological disorder, in which brain activity becomes abnormal, leading to sensations and sometimes impaired consciousness. It is characterized by recurrent, unprovoked seizures. A seizure is a transient symptom of synchronous neuronal activity in the brain\cite{Fisher}. In the world, more than 50 million people suffer from epilepsy\cite{Megiddo}. The patients with epilepsy are subjected to lifestyle limitations, such as acquiring and using a driving licence, and the social stigmatisation that often accompanies epilepsy\cite{Elger}. In current clinical practices, electroencephalography (EEG), which records the electrical activities generated by neurons in the brain, is commonly used for the epilepsy diagnosis. Long-term EEG records are reviewed and analyzed by well-trained neurologists in order to identify the occurrence of seizures and localize their zones in the brain. The seizure information is critical for physicians making therapeutical decisions, especially before any surgical operation in the cortex. However, this manual way of reviewing and analyzing is labor-intensive and time-consuming, because it usually takes several hours for a well-trained neurologist to analyze one-day recordings from one patient\cite{Gotman1982,Thodoroff,Furbass,Zandi,Shoeb2010,Shoeb2009}. This limitation motivates research of automatic seizure detection. In this paper, we will focus on developing an automatic approach to classify seizure segments and nonseizure segments from off-line EEG data records. The automatically classified seizure segments are provided to neurologists to make further analyses.

One major problem in the seizure/nonseizure classification is the significant variations of EEG signals for seizure states and nonseizure states across individuals. Also, it is a problem that the signal properties of seizures may resemble the characteristics of normal EEG signals\cite{Kiranyaz}. Furthermore, EEG signals usually contain physiologic artifacts from involuntary body or organ movements and non-physiologic noise\cite{Kiranyaz}. Machine learning techniques and signal processing technologies have been applied to address the problems. Patient-specific detectors are developed to detect seizure onsets\cite{Zandi,Shoeb2010,Amin,Hunyadi,Esbroeck,Vidyaratne}. These studies convert the problem of seizure detection into the seizure/nonseizure classification problem but more of a real-time flavor. Using traditional machine learning methods, hand-crafted features are usually needed to capture characteristics of seizure manifestations in EEG. Recently, epilepsy researchers have focused on developing seizure detection approaches based on deep learning techniques\cite{Thodoroff,Vidyaratne,Truong2018,Golmohammadi,Acharya,Kiranyaz}. Most of these deep learning-based approaches are developed based on classical neural network models, like convolutional neural network (CNN), recurrent neural network (RNN), long short-term memory (LSTM) and gated recurrent unit (GRU).
Their architectures are shallow. And they process data on different channels in the same way. A shallow neural network usually has limited capability of extracting seizure features. Different brain regions have different contributions to seizures. EEG data from different brain regions need be differentiated. Thus, we center on developing a deep neural network, which can directly process raw EEG data, differentiate data on different channels, and automatically classify seizures and nonseizures.

EEG data are one-dimensional, dynamic and non-linear\cite{Andrzejak}. RNN is better at processing one dimensional sequence data than CNN. However, RNN usually suffers from the gradient vanishing or exploding problem, and its two variants, i.e., LSTM and GRU, do not effectively support stacking multiple layers because of the gradient decay over layers\cite{Shuai_Li}. An emerging variant of RNN, independently recurrent neural network (IndRNN), addresses the above limitations. By taking the Hadamard product (i.e., element-wise product) over the recurrent inputs\cite{Shuai_Li}, it overcomes the gradient vanishing or exploding problems, and supports computations over multiple layers efficiently. Multiple layers help to address considerable variations of seizure morphologies. IndRNN is also able to process longer sequence data than LSTM. On the other hand, as information passes through many layers in a deep neural network architecture, it may possibly decay. A densely connected structure, in brief, dense structure, in which one layer connects with all preceding layers, can help preserve maximum information flow between layers\cite{Gao-Huang}. Additionally, EEG signals from different brain regions have different strengths of signifying seizures. Based on the above observations and analyses, we integrate IndRNN with the dense structure and an attention mechanism to design a deep learning approach with multiple layers for the seizure/nonseizure classification. Firstly, an attention layer is designed to differentiate data from different brain regions. It adaptively generates weights on channels and outputs weighted data. A group of IndRNN layers and batch normalization layers are organized according to the dense structure, and they constitute a dense block. Batch normalization layers are used to help reduce the over-fitting risk in deep neural network. The dense block extracts temporal spatial features from the weighted data. Predominant features at a specific temporal scale are extracted from the outputs of a dense block by deploying a max-pooling layer. After several runs of feature computations in dense blocks and max-pooling layers, the overall features within the whole temporal duration are calculated by an average pooling layer. Lastly, two fully connected (FC) layers are deployed for further integrating features and for final classification. We perform cross-validation (CV) experiments over the noisy EEG data set of CHB-MIT to evaluate our proposed approach. In our experiments, we obtain the average sensitivity, specificity and precision of 88.80\%, 88.60\% and 88.69\%, respectively, and the corresponding standard deviations of 0.0252, 0.0250, and 0.0215, respectively. The results exceed the performance of current state-of-art approaches \cite{Acharya} \cite{Hussein}. Besides, we explore how the segment length affects the performance of seizure/nonseizure classification. Our experimental results with different segment lengths show that the performance of seizure classification fluctuates over the segment lengths, and the maximal fluctuating margin is more than 4\%.

The main contributions of our paper include the following:
\begin{itemize}
\item[i)] An emerging deep learning model, IndRNN, is applied to seizure/nonseizure classification for the first time, and temporal spatial features are extracted with a deep architecture;
\item[ii)] Dense structure and attention mechanism are integrated with IndRNN for a deep neural network, and they help improve the capability of discriminating seizures from nonseizures;
\item[iii)] The relationship between the segment length and the performance of seizure/nonseizure classification is investigated.
\end{itemize}

The rest of this paper is organized as follows. Section \ref{section-related-work} describes related research about automatic seizure/nonseizure classification. Section \ref{section-method} illustrates our proposed approach of dense IndRNN with attention. The proposed approach is evaluated in CV experiments in Section \ref{section-evaluation}. Section \ref{section-model-analyses} explains the attention mechanism and validates main modules in our designed approach. In Section \ref{section-effects-segment-length-on-classification}, how the segment lengths affect the performance of classifying seizure/nonseizure is explored. Section \ref{section-discussion} discussed the approach of dense IndRNN with attention. And conclusions and future work are described in Section \ref{section-conclusion}.

\section{Related Work}
\label{section-related-work}
Seizure/nonseizure classification distinguishes seizure segments from nonseizure segments, which can be used to detect whether a data segment contains a seizure or not. For this task, extensive studies have been performed. Because seizure detection, which is often of a real-time flavor, is often treated as the seizure/nonseizure classification problem, many machine learning methods have been developed \cite{Shoeb2010,Amin,Hunyadi,Esbroeck,Vidyaratne,Fergus,Nicalaou,Kharbouch,Bolagh}. Recently, deep learning techniques have been applied to the seizure detection problem\cite{Thodoroff,Truong2018,Golmohammadi,Acharya,Hussein,Kiranyaz,Ansari}.

Esbroeck et al. developed a multi-task learning framework to detect patient-specific seizure onset\cite{Esbroeck}. In the framework, distinguishing the windows of each seizure from nonseizure data was treated as a separate task, and discriminating individual-seizures as another task. Evaluation results over the CHB-MIT data set indicated that the framework performed better in most cases compared to the standard SVM.

Kiranyaz et al. presented a systematic approach for patient-specific classification of long-term EEG\cite{Kiranyaz}. The approach can be divided into three main steps. The first step is to process data through band-pass filtering, feature extraction, epileptic seizures aggregation and morphologic filtering. The second step is to classify signal from each channel of the processed data by using a collective network. The third step is to integrate the initial classification results over each channel and make final classification decision for each EEG data segment. Over the data set of CHB-MIT, the obtained average sensitivity and specificity were 89.01\% and 94.71\%, respectively. In the approach, the many classifiers increased the computational complexity.

Based on signal processing techniques, Zandi et al. decomposed EEGs from each channel by wavelet packet transform, and separated the seizure and nonseizure states by developing a patient-specific measure\cite{Zandi}. Using the measure, a combined seizure index was derived for each epoch of every EEG channel. The combined seizure index was inspected and it triggered alarms.

Acharya et al. proposed a method to automatically detect the normal, pre-ictal, and ictal conditions from EEG signals\cite{Acharya2012}. The method firstly extracted four entropy features, including approximate entropy, sample entropy and two phase entropies, then fed the four features to classifier to do classification.

Zhou et al. designed a seizure detection algorithm based on lacunarity and Bayesian linear discriminant analysis (BLDA)\cite{Zhou}. The critical step in the algorithm was feature extraction. Firstly, EEGs were performed wavelet decomposition with five scales, and the wavelet coefficients at scales 3, 4, and 5 were selected. At the three scales, features including lacunarity and fluctuation index were extracted. Then they were passed on to the BLDA for training and classification. Over intracranial EEG data from the Epilepsy Center of the University Hospital of Freiburg, the obtained average sensitivity was 96.25\%, with an average false detection rate of 0.13 per hour and a mean delay time of 13.8s. The obtained precision results for eleven patients were less than 50\%.

Fan and Chou leveraged a complex network model to represent EEG signals, and integrated it with spectral graph theory to extract spectral graph theoretic features for detecting seizure onsets in real-time\cite{Fan}. The method was evaluated over the CHB-MIT data set. The resulting patient-specific average sensitivity was 98\%, and the latency was 6s.

The methods developed in \cite{Zandi}\cite{Kiranyaz}\cite{Esbroeck}\cite{Acharya2012}\cite{Zhou}\cite{Fan} are mostly based on signal processing methods and traditional machine learning methods. They often need crafted features, which may not be optimal.

Shoeb and Guttag leveraged the support vector machine (SVM) to develop a patient-specific detection method\cite{Shoeb2010}. Filters were applied to extract spectral features over channels. The feature vectors were concatenated according to a fixed time length and then taken as inputs to train the SVM model. The method achieved a sensitivity of 96\%, a median detection delay of 3 seconds and a median false detection rate of 2 per 24 hour. The results are often used as a benchmark for patient-specific seizure detection on the data set CHB-MIT. The authors observed that the identity of channels could help differentiate between the seizure and the nonseizure activity.

Amin and Kamboh designed an algorithm RUSBoost to process imbalanced seizure/nonseizure data, and combined RUSBoost with the decision tree classifier to do patient-specific seizure detection\cite{Amin}. The algorithm extracts spectral, spatial and temporal features from each channel. Over the CHB-MIT data set, the obtained average accuracy, sensitivity and false detection rate were 97\%, 88\%, and 0.08 per hour, respectively. Using the method, the training was fast.

Hunyadi et al. presented a nuclear norm regularization to convey multichannel information of ictal patterns\cite{Hunyadi}. The regularization integrating with extracted features helped reach a median sensitivity of 100\%, false detection rate of 0.11 per hour and alarm delay of 7.8s over the CHB-MIT data set. The proposed method processed spatial information in the same way.

In \cite{Fergus}, Fergus et al. developed a seizure detection method based on selected features from special brain regions. Over the data set of CHB-MIT, an average sensitivity of 88\% and specificity of 88\% were achieved.

Truong et al. designed an automatic seizure detection method, in which one important step is to select channels that contribute the most to seizures\cite{Truong2017}. Features, such as spectral power and correlations between channel pairs, were extracted. The classifier of Random Forest was used for classification. Over an intracranial electroencephalography (iEEG) data set, the method reached the state-of-the-art computational efficiency while maintaining the accuracy. In the method, the step of selecting channels was to reduce the number of channels, thereby improving the computational efficiency.

The approaches developed in \cite{Shoeb2010}\cite{Amin}\cite{Hunyadi}\cite{Fergus}\cite{Truong2017} extracted multichannel information to do seizure detection. However, they did not differentiate channels.

Thodoroff et al. presented a recurrent convolutional neural network to capture spectral, spatial and temporal features of
seizures\cite{Thodoroff}. EEG signals were firstly transformed into images. Created images were fed to CNN. Output vectors of the CNN were organized to be sequences in the chronological order. The sequences were passed on to the bidirectional RNN to make classification. Both patient-specific experiments and cross-patient experiments were performed. In the cross-patient testing, the obtained average sensitivity was 85\% with the false positive rate of 0.8 per hour. Transfer learning technique was utilized to overcome the problem of small amount of data in the patient-specific experiments.

Vidyaratne et al. proposed a deep recurrent architecture by combining cellular neural network with bidirectional RNN\cite{Vidyaratne}. The bidirectional RNN was deployed into each cell of the cellular neural network to extract temporal features in the forward and the backward directions. Each cell interacts with its neighboring cells to extract local spatial-temporal features. The method was evaluated in patient-specific experiments over five patients from CHB-MIT. The obtained sensitivities are 100\%. The experiments were limited, because that only five patients were tested.

Golmohammadi et al. explored two kinds of neural networks over TUH EEG Corpus\cite{Golmohammadi}. Their experimental results showed that convolutional LSTM network outperformed convolutional GRU network. Different initialization and regularization methods were also tested.

Hussein et al. designed a deep neural network for seizure/nonseizure classification by using LSTM\cite{Hussein}. The neural network extracted temporal features by using LSTM. Acharya et al. presented a 13-layers deep neural network for seizure/nonseizure classification by using CNN\cite{Acharya}. The two approaches were evaluated over the same EEG data set provided by University of Bonn\cite{Andrzejak}. The LSTM approach achieved performances of 100\%. The CNN approach obtained a sensitivity of 95\% and specificity of 90\%. Each record in the Bonn EEG data set contains only one channel and has no artifacts. And thus the Bonn data set is regarded as an easy data set.

Ansari et al. aimed to automatically optimize feature selection for seizure detection\cite{Ansari}. They utilized deep CNN to extract optimal features, and then fed the features to random forest to do classification. In evaluation experiments, EEG recordings of 26 and 22 neonates were taken as training data and testing data, respectively. A false alarm rate of 0.9 per hour and a sensitivity of 77\% were achieved. The proposed method needed no predefined features, and surpassed three classic feature-based approaches.

The deep learning approaches developed in \cite{Thodoroff}\cite{Vidyaratne}\cite{Golmohammadi}\cite{Acharya}\cite{Hussein}\cite{Ansari} are based on classic neural network models, including CNN, RNN, LSTM, and GRU. The four models have limitations for processing EEG data. CNN is good at processing two or more dimensional data but not suited to one dimensional sequence data. RNN suffers from the gradient vanishing or exploding problem. Even though LSTM and GRU improved on RNN, the two variants still have the gradient decay problem when multiple layers are deployed. We aim to build a deep neural network architecture to overcome the variability of seizure morphologies well.

\section{Methods}
\label{section-method}
\subsection{Model Design}
EEG signals are dynamic one-dimensional data. The data at one time point in an EEG signal correlate with the past data. RNN and its variants are good at processing such kind of one-dimensional data. Additionally, EEG signals for seizure and nonseizure states across individuals manifest considerable variations, and the signal properties of one patient's seizure may closely resemble the characteristics of a normal EEG signal from the same patient or other patients\cite{Kiranyaz}. In deep learning methods for classification, deep architecture is generally helpful to extract discriminative features. A variant of RNN, IndRNN, supports stacking multiple layers\cite{Shuai_Li}. And it can address the gradient vanishing and exploding problems, which restrict the effectiveness of RNN. The two variants, LSTM and GRU, suffer from the gradient decaying problem over layers. IndRNN can also process longer sequences than LSTM\cite{Shuai_Li}. Therefore, we choose to use IndRNN as the main module to construct a deep architecture for classifying seizures against nonseizures.

In a deep neural network, as information about inputs or extracted features passes through many layers, it may vanish by the time it reaches the end of the network\cite{Gao-Huang}. A densely connected structure can maximize information flow between layers in a network\cite{Gao-Huang}. We will leverage a densely connected structure to inter-connect IndRNN layers to preserve information.

A brain activity state is jointly described by EEG signals from different brain regions. Different brain regions have different contributions to a seizure state. The EEG signals at different brain regions can be differentiated when identifying seizures and nonseizures. We will introduce an attention mechanism to generate weights for channels. The weight describes the importance of EEG signal on a channel in discriminating seizures against nonseizures.

Based on the above observations, we will integrate IndRNN with a dense structure and attention mechanism to design a new approach for the seizure/nonseizure classification. The inputs are firstly passed on to an attention layer, in which attention weights are generated and multiplied with signal data. The outputs of the attention layer are weighted signals. The weighted signals are fed to the first dense block. A dense block consists of multiple IndRNN layers and is to extract temporal-spatial features in signals. In a dense block, each IndRNN layer passes its outputs to all subsequent layers; the inputs of the first IndRNN layer are also fed to all the other layers; and the outputs of the last IndRNN layer is taken as the outputs of the dense block. Each dense block is followed by a max-pooling layer. The max-pooling layer extracts predominant features at a specific time scale, and its processed results are passed on to the next dense block. The outputs of the last max-pooling layer, which follows the last dense block, are fed to an average pooling layer. In the average pooling layer, the overall features over time steps are extracted. Two FC layers follow the average layer in order to extract further features and make predictions.

\subsection{Model Architecture}
\begin{figure*}[htb]
\centering
\includegraphics[width=6.5in]{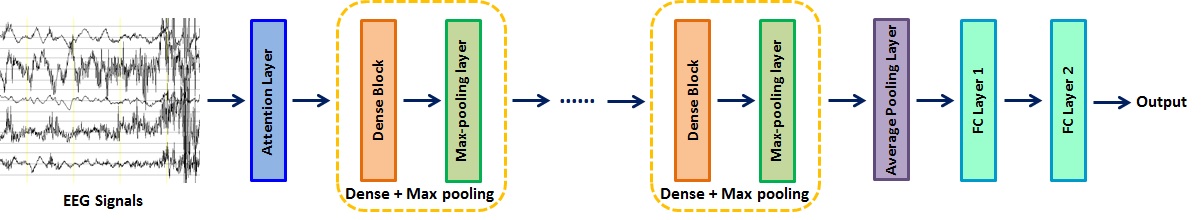}
\caption{Architecture of the proposed ADIndRNN approach.}
\label{architecture-attention-dense-IndRNN-approach}
\end{figure*}

The proposed integrated IndRNN with a dense structure and attention mechanism is called dense IndRNN with attention (ADIndRNN). It consists of an attention layer, dense blocks, max-pooling layers, an average pooling layer, and two FC layers. Each dense block further comprises of IndRNN layers and bach normalization layers. The architecture of the ADIndRNN is presented in Fig. \ref{architecture-attention-dense-IndRNN-approach}.

\subsubsection{Attention Layer}
\begin{figure}[t]
\centering
\includegraphics[width=1.55in]{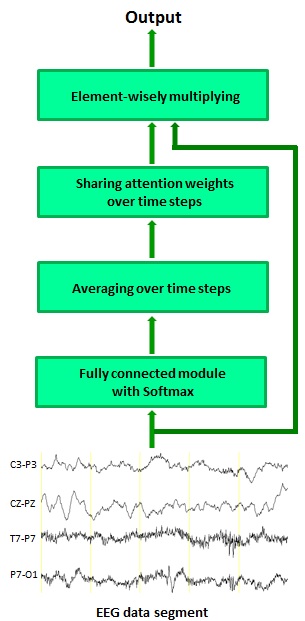}
\caption{Work flow of attention layer}
\label{workflow-attention-layer-in-ADIR}
\end{figure}

 The attention layer adaptively generates weights for channels and executes element-wise multiplication between data segments and weights. It outputs weighted signals. Its work flow is given in Fig. \ref{workflow-attention-layer-in-ADIR}, and its computation is described in  (\ref{eq-atten-layer-reshape-1})$-$(\ref{eq-atten-layer-element-wise-multi}). The attention weights are generated according to an attention mechanism. Each data segment executes one linear transformation based on a kernel matrix and a bias matrix. After the linear transformation, the $softmax(\cdot)$ function is applied to each time step separately. The activation function outputs weights on channels at each time step. In order to diminish differences of channel weights among time steps, the weights over time steps are averaged. And the averages are taken as weights on channels at these time steps. The attention mechanism is illustrated in (\ref{eq-atten-layer-reshape-1})$-$(\ref{eq-atten-layer-copy}).

    \begin{align}
    \label{eq-atten-layer-reshape-1} Y_{1}  & =  f_{re_{1}}(X_{0})  \\
    \label{eq-atten-layer-fully-connected} Y_{2} & =  softmax(Y_{1} * W_{al} + B_{al})  \\
    \label{eq-atten-layer-reshape-2} Y_{3} & = f_{re_{2}}(Y_{2})  \\
    \label{eq-atten-layer-average} Y_{4} & = f_{av}(Y_{3})  \\
    \label{eq-atten-layer-copy} Y_{5} & = f_{cy}(Y_{4})  \\
    \label{eq-atten-layer-element-wise-multi} Y_{al} & = X_{0}\odot Y_{5}
    \end{align}
    Here, $X_{0}$ denotes an input tensor of size $(n_{sm},n_{sp},n_{ch})$. Symbols $n_{sm}$, $n_{sp}$, $n_{ch}$ represent the numbers of samples, time steps, and signal channels, respectively. $Y_{1}$ is a matrix of size $(n_{ss},n_{ch})$, $n_{ss}=n_{sm}*n_{sp}$, $W_{al}$ a weight matrix of size $(n_{ch},n_{ch})$, a bias matrix $B_{al}$ of size $(n_{ss},n_{ch})$, and $Y_{2}$ with size $(n_{ss},n_{ch})$. $softmax(\cdot)$ is a normalized exponential function. $Y_{3}$ is a matrix of size $(n_{sm},n_{sp},n_{ch})$, $Y_{4}$ of size $(n_{sm},n_{ch})$, $Y_{5}$ of size $(n_{sm},n_{sp},n_{ch})$, and $Y_{al}$ an output matrix of attention layer with shape $(n_{sm},n_{sp},n_{ch})$. Functions $f_{re_{1}}(\cdot)$ and $f_{re_{2}}(\cdot)$ are to reshape a matrix, $f_{av}(\cdot)$ is a function of computing averages along with the second axis of matrix, and $f_{cy}(\cdot)$ is an copying operation to share the averages over all the time steps. The symbol $\odot$ means an element-wise multiplication between matrices.

\subsubsection{Dense Block}
\begin{figure*}[htb]
\centering
\includegraphics[width=5.7in]{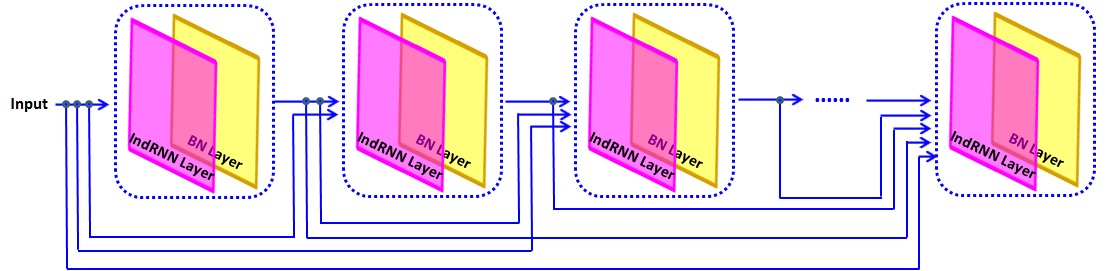}
\caption{Architecture of dense block.}
\label{architecture-dense-block}
\end{figure*}

A dense block organizes its components in a densely connected fashion to ensure maximum information flow between layers. Each component consists of one IndRNN layer and one batch normalization (BN) layer, and the IndRNN layer is followed by the BN layer. Outputs of one component are passed on to all subsequent components. And the inputs of the first component are fed to other each component. The structure of a dense block is shown in Fig. \ref{architecture-dense-block}.

    An IndRNN layer processes input sequences in forward order, and extracts time-dependent features. Its computation is as follows:
    \begin{align}
        \label{eq-hidden-layer-IndRNN} H_{t} & =  \sigma_{hid}(X^{IR}_{t} * W_{hid} + H_{t-1} \odot U_{hid} + B_{hid}) \\
        \label{eq-output-IndRNN} Y^{IR}_{t} & = \sigma_{out}(H_{t} * W_{out} + B_{out})
    \end{align}
    Here, $X^{IR}_{t}$ is a matrix of size $(n_{sm}, n_{fe})$, which represents an input of IndRNN layer at the time step $t$. $n_{fe}$ is the number of features. $H_{t}$ is a matrix of size $(n_{sm}, n_{hid})$, and it means hidden outputs at the time step of $t$ in IndRNN layer. $n_{hid}$ represents the number of hidden states. $W_{hid}$ is a input weight matrix with shape $(n_{fe}, n_{n_{hid}})$, $U_{hid}$ for a recurrent weight matrix of size $(n_{sm}, n_{hid})$. $B_{hid}$ and $B_{out}$ are two bias matrices of size $(n_{sm}, n_{hid})$. $W_{out}$ is an output weight matrix of size $(n_{hid}, n_{hid})$. $Y^{IR}_{t}$ is a matrix of size $(n_{sm}, n_{hid})$, which means an output at the time step $t$ in the IndRNN layer. $\sigma_{hid}$ and $\sigma_{out}$ are activation functions such as ReLU.

    For the first IndRNN layer in the first dense block, its input at the time step $t$ is the output of the attention layer at the time step $t$, i.e., $X^{IR}_{t}$ consists of all the elements at the entry of $t$ along the axis of 1 in the matrix $Y_{al}$.

    BN layer is inserted after each IndRNN layer. It is used to speed up training and reduce overfitting\cite{Sergey_Ioffe}.

\subsubsection{Max-pooling Layer}
Each dense block is followed by a max-pooling layer. The max-pooling layer extracts predominant features from output sequences of dense block at a specific temporal scale.

\subsubsection{Average Pooling Layer}
Average pooling layer is designed to extract overall features across time scales for the final classification. Its output is a two-dimensional matrix, in which each row corresponds to one segment sample and elements in columns are features. It is inserted after the last max-pooling layer.

\subsubsection{FC Layers}
Two FC layers are deployed behind the average pooling layer. The first FC layer aims to integrate features from the outputs of the average pooling layer and make further extractions. The second is to perform final classification of seizure/nonseizure.

\section{Evaluation}
\label{section-evaluation}
In order to evaluate our proposed approach of ADIndRNN, we conduct CV experiments over the noisy EEG data set of CHB-MIT, and measure its performance in five metrics. The five metrics are sensitivity, specificity, F1 score, precision and accuracy, respectively. The proposed approach is compared with two current state-of-the-art approaches, including one LSTM-based approach\cite{Hussein} and one CNN-based approach\cite{Acharya}. The CV is that, data segments from all the patients are put into a pool, then the segments in the pool are randomly split into three disjoint sets according to a ratio, including training dada set, validation data set and testing data set. The training data are to train a model, the validation data are to tune parameters in the model based on validation performance, and the testing data are to test the generalization capability of trained model. To reduce the variability of testing results, ten rounds of CVs are performed for each approach, then averages and standard deviations for the ten CV results are calculated as the performance of seizure/nonseizure classification approach.

\subsection{Data Set}
The data set of CHB-MIT\cite{Shoeb2009} contains 686 EEG recordings from 23 subjects of different ages ranging from 1.5 years to 22 years. The recordings include 198 seizures. The sampling frequency is 256 Hz. Most recordings are one hour long, and others are two-hour long or four-hour long. The EEG recordings are grouped into 24 cases. In each case, the data recordings are from a single subject. Case Chb21 was obtained 1.5 years after Case Chb01 from the same subject. Each data file contains data on 23 or more channels. There exist channels on which data are missing. Thus, we only consider those channels containing all the original data. Three data files, including Chb12\_27.edf, Chb12\_28.edf, and Chb12\_29.edf, have different channel montages from other data files. In our experiments, we remove these three data files.

\subsection{Data Segmentation}
\label{section_data_segmentation}
In order to extract effective seizure features, 17 common channels are chosen. According to a segment length of 23 seconds, each data record in each case is split into data segments from the beginning to the end. Data segments except for the last one of a data record have no overlap. If the duration of a data record is not divided by the segment length and there is a seizure happening in the remaining part, we will ensure that the last segment will have the same length but will overlap with its prior segment. If the remaindering part contains no seizure, then it is dropped. Using annotation files for this data set, we determine whether a data segment contains a seizure or not. In our experiments, if a segment contains any seizure data, it is considered as a seizure segment; otherwise, it is a nonseizure segment.

Using the above segmentation method, 665 seizure segments are obtained. In these seizure segments, the lengths of seizures vary from 1s to 23s, with the average length being 16.9s. Among all seizure segments, segments containing seizure data of less than 7s comprise 14.7\%, those containing more than 17s comprise 59.8\%, and those containing more than 10s comprise 76.1\%. All the seizure segments are taken as a part of our experiment data. We randomly choose 665 nonseizure segments in each experiment. The 1330 seizure/nonseizure segments are randomly split into training set, validation set and testing set with a ratio of 70:15:15 in each experiment, and we adopt the repeated random sub-sampling validation as a strategy for CV.

\subsection{CV Results for the Proposed Approach}
Based on the architecture in Fig. \ref{architecture-attention-dense-IndRNN-approach}, we build a model by stacking three dense blocks, each dense block consisting of three IndRNN layers and three BN layers. The constructed model is denoted by ADIndRNN-(3,3). In our CV experiments using the model ADIndRNN-(3,3), the total loss contains two parts. One part is losses produced by predicted labels of the model and ground truth labels. The other part is L2 losses of all the trainable variables. For the second loss part, we set a coefficient of weight decay to take its specific percentage into the total loss. The main parameters in the model are set as follows: A kernel initializer in the attention layer is a truncated normal initializer with mean value of 0 and standard deviation of 0.1, a bias initializer in the attention layer initializes  tensor to 0; the state size of each IndRNN layer in the first dense block is 80, that in the second dense block is 120, that in the third dense block is 160; each max-pooling layer has a window size of 2 and stride of 2; weight initializers in the two FC layers are Xavier initializers, two bias initializers in the two FC layers are constant initializers with value of 0.001, the number of output units in the two FC layers are 100 and 2, respectively; the weight decay for the trainable variables loss is 0.01; the optimizer is Adam, the learning rate is 0.0004; the batch size for training is 30, the epochs is 60. Overall, ten rounds of CVs are performed. The obtained results of these ten experiments over 23s segments using ADIndRNN-(3,3) are given in Table \ref{cross-validation-results-for-proposed-approach}. The averages (abbrev. Ave.) in Sensitivity (abbrev. Sens.), Specificity (abbrev. Spec.), F1 score, Precision (abbrev. Prec.) and Accuracy (abbrev. Acc.) are 88.80\%, 88.60\%, 88.71\%, 88.69\%, and 88.70\%, respectively; and the standard deviations (abbrev. Std.) are 0.0252, 0.0250, 0.0134, 0.0215, and 0.0133, respectively.

\begin{table}[ht]
\centering
\caption{CV results using the proposed approach}
\label{cross-validation-results-for-proposed-approach}
\begin{tabular}{|c||c|c|c|c|c|}
\hline
\textbf{Item} & \textbf{Sens.} & \textbf{Spec.} & \textbf{F1 Score} & \textbf{Prec.} & \textbf{Acc.} \\
\hline
1 & 0.9100 & 0.8500 & 0.8835 & 0.8585 & 0.8800 \\
\hline
2 & 0.8600 & 0.9300 & 0.8912 & 0.9247 & 0.8950  \\
\hline
3 & 0.9100 & 0.8700 & 0.8922 & 0.8750 & 0.8900 \\
\hline
4 & 0.8500 & 0.9200 & 0.8808 & 0.9140 & 0.8850 \\
\hline
5 & 0.8600 & 0.8600 & 0.8600 & 0.8600 & 0.8600 \\
\hline
6 & 0.9000 & 0.8900 & 0.8955 & 0.8911 & 0.8950 \\
\hline
7 & 0.9300 & 0.8800 & 0.9073 & 0.8857 & 0.9050 \\
\hline
8 & 0.8700 & 0.8800 & 0.8744 & 0.8788 & 0.8750 \\
\hline
9 & 0.8900 & 0.8700 & 0.8812 & 0.8725 & 0.8800 \\
\hline
10 & 0.9000 & 0.9100 & 0.9045 & 0.9091 & 0.9050 \\
\hline
\hline
\textbf{Ave.} & \textbf{0.8880} & \textbf{0.8860} & \textbf{0.8871} & \textbf{0.8869} & \textbf{0.8870} \\
\hline
\textbf{Std.} & \textbf{0.0252} & \textbf{0.0250} & \textbf{0.0134} & \textbf{0.0215} & \textbf{0.0133} \\
\hline

\end{tabular}
\end{table}

\subsection{Comparison with the LSTM and CNN Approaches}

LSTM as a main module has been used to detect seizures\cite{Hussein}. The LSTM approach is evaluated through CV experiments over an EEG data set from University of Bonn\cite{Andrzejak}, showing state-of-the-art performance. A CNN-based approach has been proposed for seizure/nonseizure classification\cite{Acharya}, which also demonstrates state-of-the-art performance over the Bonn data set. Because the Bonn data set is heavily processed and contains no artifacts, and its size is small, we compare the proposed approach with the LSTM approach and the CNN approach over the noisy EEG data set of CHB-MIT.

For the LSTM approach and the CNN approach, we implement them according to the descriptions in the related literature. And the implementations are tested. Our obtained testing results reach the reported performances. Using the two implementations, CV experiments are performed for the LSTM approach and the CNN approach separately. The LSTM approach consists of one LSTM layer, one time-distributed computing layer, one average pooling layer and one FC layer.
In the experiments using the LSTM approach, our parameter setting is as follows: The number of hidden states is 120 in the LSTM layer, that in the time-distributed computing layer is 60, the optimizer is RMSprop, the learning rate is 0.0007, the batch size is 30, and the number of epochs is 30.
For the CNN approach, it contains five convolutional layers, five max pooling layers, and three FC layers. The parameters are set as follows: the number of hidden states in the first two convolutional layers is 100, that in each of the second two convolutional layers is 200, that in the fifth convolutional layer is 260, that in the first FC layer is 100, that in the second FC layer is 50, the parameter alpha is 0.01 in the LeakyReLU activation function, the optimizer is Adam, the learning rate is 0.001, the batch size is 30, and the number of epochs is 50. Using the LSTM approach, ten rounds of CVs over 23s segments from CHB-MIT are performed, and their obtained results are shown in Table \ref{cross-validation-results-LSTM}. The obtained average sensitivity is 84.40\%, the average specificity is 84.30\%, and the average precision is 84.70\%. For the CNN approach, ten rounds of CV experiments over 23s segments are also conducted, and the results are given in Table \ref{cross-validation-results-CNN}. The achieved average sensitivity, the average specificity and the average precision are 84.80\%, 81.00\%, and 82.56\%, respectively.

\begin{table}[ht]
\centering
\caption{CV results using the LSTM approach}
\label{cross-validation-results-LSTM}
\begin{tabular}{|c||c|c|c|c|c|}
\hline
\textbf{Item} & \textbf{Sens.} & \textbf{Spec.} & \textbf{F1 Score} & \textbf{Prec.} & \textbf{Acc.} \\
\hline
1 & 0.8500 & 0.8800 & 0.8629 & 0.8763 & 0.8650 \\
\hline
2 & 0.7700 & 0.8500 & 0.8021 & 0.8370 & 0.8100 \\
\hline
3 & 0.7900 & 0.8700 & 0.8229 & 0.8587 & 0.8300 \\
\hline
4 & 0.7100 & 0.9300 & 0.7978 & 0.9103 & 0.8200 \\
\hline
5 & 0.8200 & 0.8900 & 0.8497 & 0.8817 & 0.8550 \\
\hline
6 & 0.9100 & 0.7900 & 0.8585 & 0.8125 & 0.8500 \\
\hline
7 & 0.8600 & 0.8300 & 0.8473 & 0.8350 & 0.8450 \\
\hline
8 & 0.8600 & 0.8400 & 0.8515 & 0.8431 & 0.8500 \\
\hline
9 & 0.9400 & 0.7200 & 0.8468 & 0.7705 & 0.8300  \\
\hline
10 & 0.9300 & 0.8300 & 0.8857 & 0.8455 & 0.8800 \\
\hline
\hline
Ave. & 0.8440 & 0.8430 & 0.8425 & 0.8470 & 0.8435  \\
\hline
Std. & 0.0696 & 0.0550 & 0.0259 & 0.0368 & 0.0201  \\
\hline

\end{tabular}
\end{table}

\begin{table}[ht]
\centering
\caption{CV results using the CNN approach}
\label{cross-validation-results-CNN}
\begin{tabular}{|c||c|c|c|c|c|}
\hline
\textbf{Item} & \textbf{Sens.} & \textbf{Spec.} & \textbf{F1 Score} & \textbf{Prec.} & \textbf{Acc.} \\
\hline
1 & 0.8400 & 0.8500 & 0.8442 & 0.8485 & 0.8450 \\
\hline
2 & 0.9200 & 0.7700 & 0.8558 & 0.8000 & 0.8450 \\
\hline
3 & 0.8000 & 0.8400 & 0.8163 & 0.8333 & 0.8200  \\
\hline
4 & 0.9000 & 0.6900 & 0.8145 & 0.7438 & 0.7950  \\
\hline
5 & 0.9200 & 0.8000 & 0.8679 & 0.8214 & 0.8600  \\
\hline
6 & 0.7900 & 0.8500 & 0.8144 & 0.8404 & 0.8200  \\
\hline
7 & 0.6300 & 0.9700 & 0.7590 & 0.9545 & 0.8000 \\
\hline
8 & 0.8500 & 0.8700 & 0.8586 & 0.8673 & 0.8600 \\
\hline
9 & 0.8700 & 0.7700 & 0.8286 & 0.7909 & 0.8200 \\
\hline
10 & 0.9600 & 0.6900 & 0.8458 & 0.7559 & 0.8250 \\
\hline
\hline
Ave. & 0.8480 & 0.8100 & 0.8305 & 0.8256 & 0.8290 \\
\hline
Std. & 0.0891 & 0.0809 & 0.0301 & 0.0571 & 0.0217 \\
\hline

\end{tabular}
\end{table}

Comparing CV results in Tables \ref{cross-validation-results-for-proposed-approach}-\ref{cross-validation-results-CNN}, we can conclude that the average performance, in either one of the metrics of sensitivity, specificity, F1 score, precision, and accuracy,  using the proposed approach is at least 4\% greater than that using the LSTM approach or the CNN approach, and the obtained standard deviation over each metric using our approach is smaller. The above comparisons show that the proposed approach outperforms the LSTM approach and the CNN approach in seizure/nonseizure classification.

\section{Model Analyses}
\label{section-model-analyses}
Our designed architecture in Fig. \ref{architecture-attention-dense-IndRNN-approach} contains the attention layer, the dense structure, and the IndRNN layer.
In this section, we will explain attention weights by taking examples of segments in CHB-MIT, validate separately the attention layer and the dense structure, and evaluate models with more IndRNN layers.

\subsection{Interpretations of Attention Mechanism}
\begin{figure}[htbp]
\centering
\includegraphics[width=3.2in]{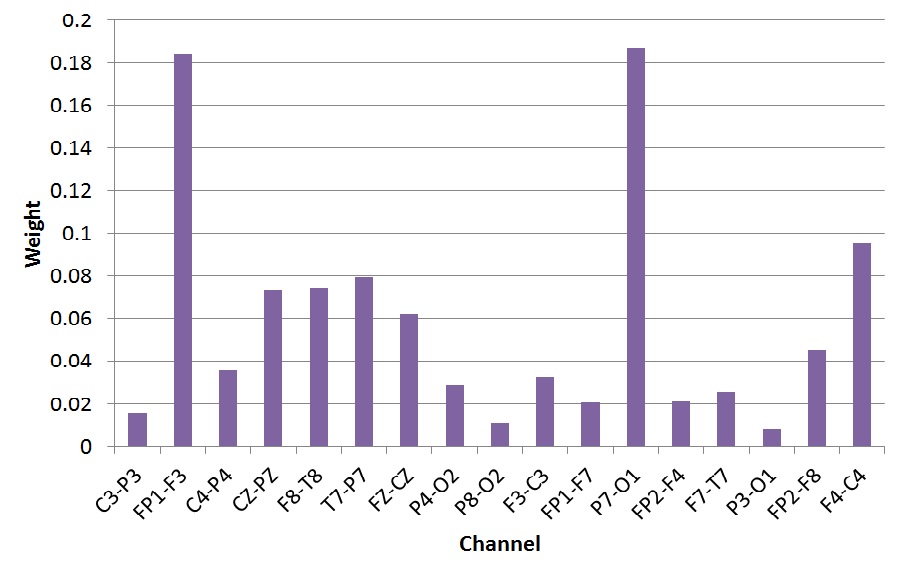}
\caption{Attention weights of one seizure segment from Chb04}
\label{atten-weight-sei-seg-13-chb04-08-model-1}
\end{figure}

\begin{figure}[htbp]
\centering
\includegraphics[width=3.2in]{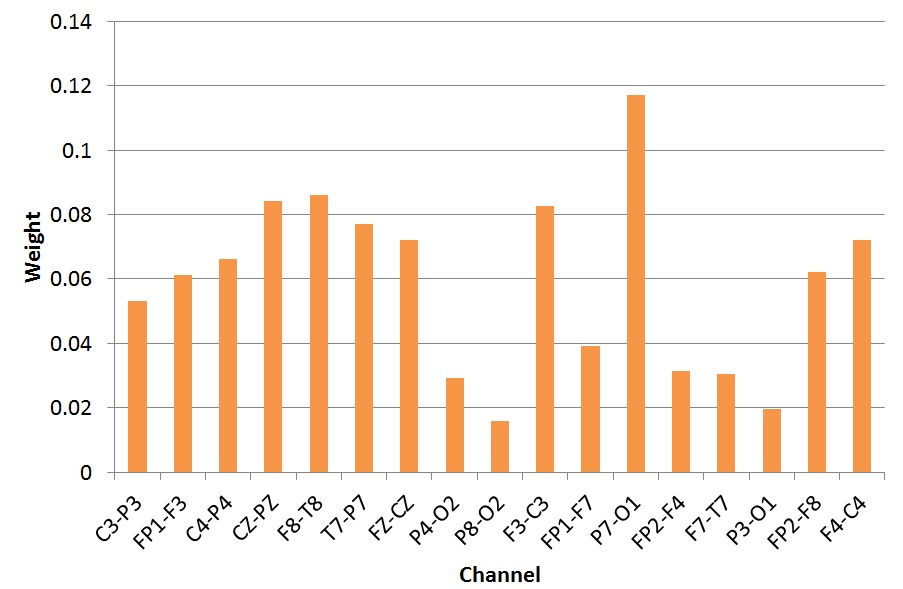}
\caption{Attention weights of one seizure segment from Chb10}
\label{atten-weight-sei-seg-17-chb10-89-model-1}
\end{figure}

Our attention mechanism differentiates information on different channels, and assigns different weights to them. The attention mechanism contains a kernel matrix and a bias matrix, which are trained together with the whole model of ADIndRNN. A data segment is combined with the two trained matrices to do multiplying and adding operations. Then, the $softmax(\cdot)$ function and an averaging operation are applied to calculate weights on channels. The obtained attention weights depend on the trained kernel matrix, the trained bias matrix, and data segment. Based on a well-trained model ADIndRNN-(3,3), Fig. \ref{atten-weight-sei-seg-13-chb04-08-model-1} shows attention weights over channels for a 23s seizure segment from Chb04, and Fig. \ref{atten-weight-sei-seg-17-chb10-89-model-1} presents weights over channels for a 23s seizure segment from Chb10. The weights in Fig. \ref{atten-weight-sei-seg-13-chb04-08-model-1} and Fig. \ref{atten-weight-sei-seg-17-chb10-89-model-1} are based on the same well-trained kernel matrix and bias matrix, but on different data segments. The two weights distributions in the two figures are different, and their differences indicate that the attention mechanism can adaptively calculate weights over channels. The magnitude of an attention weight manifests the size of differences between seizure signals and nonseizure signals on a channel. The larger an attention weight is, the better the signals on the corresponding channel can distinguish seizures from nonseizures.

\begin{figure}[h]
\centering
\subfigure[]{
\label{fig-nonsei-seg-chb04-08-model-1}
\includegraphics[width=3.1in]{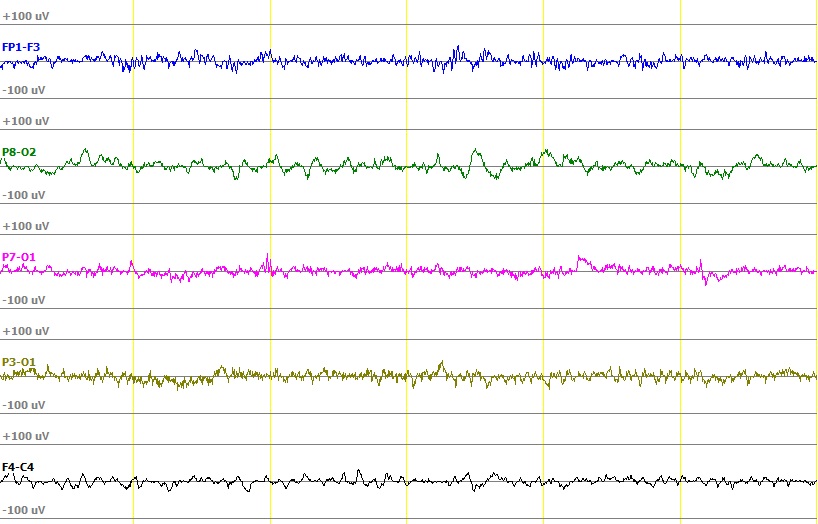}
}

\subfigure[]{
\label{fig-sei-seg-chb04-08-model-1}
\includegraphics[width=3.1in]{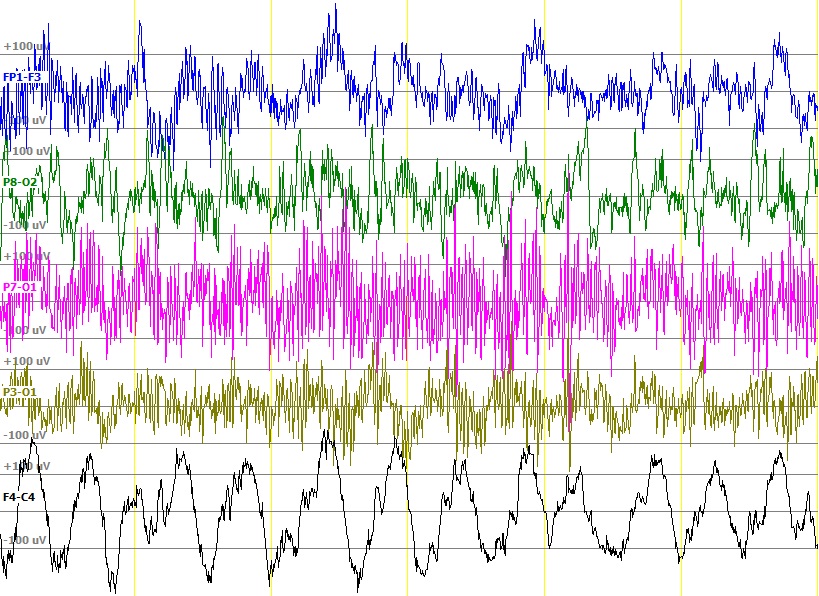}
}
%\vspace{-0.7cm}
\caption{Visualization of signals on multiple channels in segments from Chb04. (a) the signals in a nonseizure segment, (b) the signals in a seizure segment.}
\label{sei-nonsei-segments-from-chb04}
\end{figure}

Fig. \ref{fig-nonsei-seg-chb04-08-model-1} is a visualization of nonseizure signals on five channels from the patient Chb04, and Fig. \ref{fig-sei-seg-chb04-08-model-1} for seizure signals on the same five channels from the same patient. The five channels in Fig. \ref{sei-nonsei-segments-from-chb04} are FP1-F3, P8-O2, P7-O1, P3-O1, and F4-C4, respectively. The visualized seizure segment in Fig. \ref{fig-sei-seg-chb04-08-model-1} is the one whose channel weights are shown in Fig. \ref{atten-weight-sei-seg-13-chb04-08-model-1}.
In Fig. \ref{atten-weight-sei-seg-13-chb04-08-model-1}, weights on channels P7-O1 and FP1-F3 are the first two largest, and weights on channels P3-O1 and P8-O2 are the two smallest. Comparing seizure signals in Fig. \ref{fig-sei-seg-chb04-08-model-1} with nonseizure signals in Fig. \ref{fig-nonsei-seg-chb04-08-model-1}, signals on the channel P7-O1 (i.e., signals with the magenta color) change the most. The magenta seizure signal in Fig. \ref{fig-sei-seg-chb04-08-model-1} oscillates with the largest magnitude range and the largest frequency. Its magnitude ranges in five zones. The seizure signal on Channel FP1-F3 in Fig. \ref{fig-sei-seg-chb04-08-model-1} (i.e., the blue seizure signal) fluctuates more frequently than the seizure signals on channels F4-C4 and P8-O2 (i.e., the black seizure signal and the green seizure signal). The seizure signal on Channel P3-O1 (i.e., the olive green seizure signal) oscillates in a range of two zones, which is smaller than the blue seizure signal and the black seizure signal. So, the weight values in Fig. \ref{atten-weight-sei-seg-13-chb04-08-model-1} mostly characterize the sizes of differences between seizure signals and nonseizure signals in Fig. \ref{sei-nonsei-segments-from-chb04}.

\begin{figure}[h]
\addtocounter{subfigure}{2}
\centering
\subfigure[]{
\label{fig-nonsei-seg-chb10-89-model-1}
\includegraphics[width=3.1in]{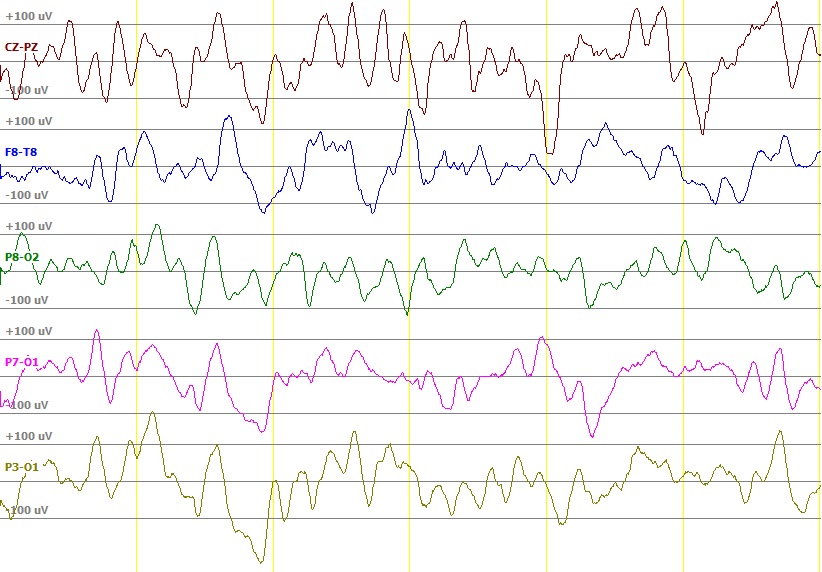}
}

\subfigure[]{
\label{fig-sei-seg-chb10-89-model-1}
\includegraphics[width=3.1in]{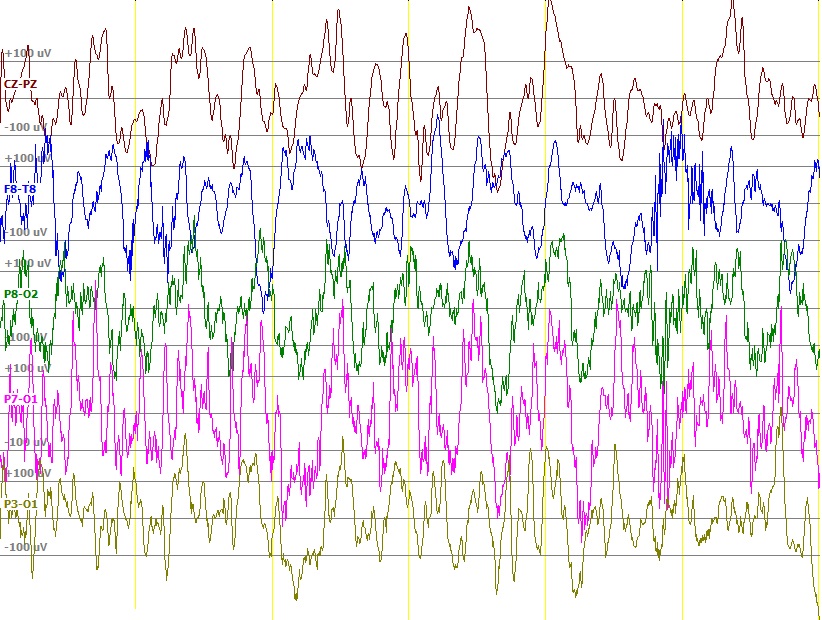}
}
%\vspace{-0.7cm}
\caption{Visualization of signals on multiple channels in segments from Chb10. (c) the signals in a nonseizure segment, (d) the signals in a seizure segment.}
\label{sei-nonsei-segments-from-chb10}
\end{figure}

Fig. \ref{fig-sei-seg-chb10-89-model-1} visualizes the seizure segment that the channel weights are presented in Fig. \ref{atten-weight-sei-seg-17-chb10-89-model-1}. Fig. \ref{sei-nonsei-segments-from-chb10} shows nonseizure signals and seizure signals on five channels from the patient Chb10. The five channels are CZ-PZ, F8-T8, P8-O2, P7-O1, and P3-O1, respectively. The seizure signal on Channel P7-O1 (i.e., the magenta seizure signal) in Fig. \ref{fig-sei-seg-chb10-89-model-1} oscillates in six zones, and fluctuates faster than the magenta nonseizure signal in Fig. \ref{fig-nonsei-seg-chb10-89-model-1}. Magnitudes of the other four seizure signals range in at the most five zones. The magenta signals change the most from a nonseizure state to a seizure state. It is just why the weight on Channel P7-O1 in Fig. \ref{atten-weight-sei-seg-17-chb10-89-model-1} is the largest. The seizure signals on channel F8-T8 and CZ-PZ (i.e., the blue seizure signal and the brown seizure signal, respectively) in Fig. \ref{fig-sei-seg-chb10-89-model-1} oscillate mostly in four zones, and the blue seizure signal fluctuates a little more frequently than the brown seizure signal. The magnitude in the seizure signal on Channel P8-O2 (i.e., the green signal) in Fig. \ref{fig-sei-seg-chb10-89-model-1} changes more slowly than the magnitude in the blue seizure signal, although the green signal oscillates a little more frequently than the blue seizure signal. The seizure signal on Channel P3-O1 (i.e., the olive seizure signal) in Fig. \ref{fig-sei-seg-chb10-89-model-1} oscillates mostly in three zones and less frequently than the blue seizure signal. So, the weight on Channel P3-O1 is smaller than weights on channels CZ-PZ and F8-T8 in Fig. \ref{atten-weight-sei-seg-17-chb10-89-model-1}.

\subsection{Validation of Modules}
\label{section-validation-of-modules}
In order to validate modules in our architecture of Fig. \ref{architecture-attention-dense-IndRNN-approach}, we take the model ADIndRNN-(3,3) as a representative and conduct CV experiments over 23s segments. The model ADIndRNN-(3,3) contains an attention layer and three dense blocks, each block including three IndRNN layers. So, we evaluate a model IndRNN-9. In this model, nine IndRNN layers are used; each IndRNN layer is followed by one BN layer and one max-pooling layer orderly; the last max-pooling layer is followed by one average pooling layer and two FC layers. A model AIndRNN-9 is constructed by inserting one attention layer before IndRNN-9 in order to test the contributions of attention mechanism. And also a model DIndRNN-(3,3), which is obtained by removing the attention layer in the model ADIndRNN-(3,3), is evaluated. The three models, including IndRNN-9, AIndRNN-9, and DIndRNN-(3,3), are separately performed ten rounds of CVs based on their tuned parameters. For each model, the averages and standard deviations over its achieved ten testing results are presented in Table \ref{experiment-results-for-modules-in-ADIndRNN-(3,3)}.

Comparing results in Table \ref{experiment-results-for-modules-in-ADIndRNN-(3,3)} with results in Table \ref{cross-validation-results-for-proposed-approach}, we can see that, using IndRNN only, or the combination of IndRNN and attention mechanism, or the combination of IndRNN and dense structure does not reach the performance of the model ADIndRNN-(3,3). After adding the attention layer, the model AIndRNN-9 is a little better than IndRNN-9, and ADIndRNN-(3,3) outperforms DIndRNN-(3,3) by at least 1\%. By using dense structure, DIndRNN-(3,3) is better than IndRNN-9, and the model ADIndRNN-(3,3) improves 4\% in the sensitivity compared with AIndRNN-9. Thus, the attention layer and the dense structure in our architecture of Fig. \ref{architecture-attention-dense-IndRNN-approach} provide positive contributions for seizure/nonseizure classification.

\begin{table*}[htb]
\centering
\caption{CV results for modules in ADIndRNN-(3,3)}
\label{experiment-results-for-modules-in-ADIndRNN-(3,3)}
  \begin{tabular}{|c|c|c|c|c|c|}
  \hline
    \textbf{Model} & \textbf{Sens.} & \textbf{Spec.} & \textbf{F1 Score} & \textbf{Prec.} & \textbf{Acc.} \\
    \hline
    IndRNN-9 & 0.8440$\pm$0.0338  & 0.8770$\pm$0.0560  & 0.8584$\pm$0.0177 & 0.8768$\pm$0.0498 & 0.8605$\pm$0.0204   \\
    \hline
    AIndRNN-9 & 0.8470$\pm$0.0300 & \textbf{0.8830$\pm$0.0326}  & 0.8625$\pm$0.0179 & \textbf{0.8798$\pm$0.0285} & 0.8650$\pm$0.0173 \\
    \hline
   DIndRNN-(3,3) & \textbf{0.8710$\pm$0.0308}  & 0.8730$\pm$0.0338  & \textbf{0.8719$\pm$0.0241} & 0.8735$\pm$0.0292 & \textbf{0.8720$\pm$0.0243} \\
    \hline
  \end{tabular}
\end{table*}

\subsection{Analyses for deeper IndRNN-related models}
The model ADIndRNN-(3,3), as a representative of the architecture in Fig. \ref{architecture-attention-dense-IndRNN-approach}, shows good performance in Table \ref{cross-validation-results-for-proposed-approach}, which is better than the LSTM approach and the CNN approach. It uses nine IndRNN layers. A natural question is what are the performances for models containing more IndRNN layers. We evaluate six models, including IndRNN-12, IndRNN-15, AIndRNN-12, AIndRNN-15, DIndRNN-(4,3), and ADIndRNN-(4,3). The models IndRNN-12 and IndRNN-15 stack layers in the similar way as IndRNN-9. Their difference is the number of IndRNN layers. IndRNN-12 uses 12 IndRNN layers, IndRNN-15 contains 15 IndRNN layers, and IndRNN-9 includes 9 IndRNN layers. The models AIndRNN-12 and AIndRNN-15 are achieved by inserting an attention layer separately before IndRNN-12 and IndRNN-15. The model ADIndRNN-(4,3) is constructed according to the architecture in Fig. \ref{architecture-attention-dense-IndRNN-approach} by using four dense blocks, each block containing three IndRNN layers. The model DIndRNN-(4,3), which contains four dense blocks with three IndRNN layers per block, is obtained by removing an attention layer from the model ADIndRNN-(4,3). Using the data segmentation method in Section \ref{section_data_segmentation}, each one of the above six models is performed for ten rounds of CV tests based on their tuning-parameters feed backs. And then the average and standard deviation of ten testing results for each model are calculated and shown in Table \ref{experiment-results-for-deeper-IndRNN-related-models}.

Compared to the results in Table \ref{experiment-results-for-modules-in-ADIndRNN-(3,3)}, the accuracy for IndRNN-12 in Table \ref{experiment-results-for-deeper-IndRNN-related-models} is a little better than that for IndRNN-9, and the accuracy for IndRNN-15 better than that for IndRNN-12; the performance of AIndRNN-12 is better than that of AIndRNN-9, and AIndRNN-15 better than AIndRNN-12. The accuracy of DIndRNN-(4,3) is close to that of DIndRNN-(3,3). The accuracy for ADIndRNN-(4,3) is a little better than that for AIndRNN-12, and equals to that of DIndRNN-(4,3). The comparisons indicate that, using more IndRNN layers can help improve the performance. However, the performances for the six models are not as good as the performance of the model ADIndRNN-(3,3). The reasons may be possibly two folds: One is that the data size of 1330 segments is not big enough to train a deeper neural network well; The other is that, the dense structure and the attention mechanism are more effective than using more layers for the seizure/nonseizure classification.

\begin{table*}[htb]
\centering
\caption{CV results for deeper IndRNN-related models}
\label{experiment-results-for-deeper-IndRNN-related-models}
  \begin{tabular}{|c|c|c|c|c|c|}
  \hline
    \textbf{Model} & \textbf{Sens.} & \textbf{Spec.} & \textbf{F1 Score} & \textbf{Prec.} & \textbf{Acc.} \\
    \hline
    IndRNN-12 & 0.8520$\pm$0.0424 & 0.8730$\pm$0.0377 & 0.8608$\pm$0.0175 & 0.8723$\pm$0.0314 & 0.8625$\pm$0.0157 \\
    \hline
    IndRNN-15 & \textbf{0.8730$\pm$0.0377} & 0.8670$\pm$0.0602 & 0.8707$\pm$0.0310 & 0.8708$\pm$0.0498 & 0.8700$\pm$0.0328  \\
    \hline
    AIndRNN-12 & 0.8560$\pm$0.0242 & 0.8810$\pm$0.0336 & 0.8669$\pm$0.0145 & 0.8792$\pm$0.0290 & 0.8685$\pm$0.0150 \\
    \hline
    AIndRNN-15 & 0.8600$\pm$0.0205 & 0.8870$\pm$0.0438 & \textbf{0.8721$\pm$0.0212} & 0.8856$\pm$0.0380 & \textbf{0.8735$\pm$0.0235} \\
    \hline
    DIndRNN-(4,3) & 0.8620$\pm$0.0399  & 0.8810$\pm$0.0318  & 0.8700$\pm$0.0179  & 0.8800$\pm$0.0255  &  0.8715$\pm$0.0160    \\
    \hline
    ADIndRNN-(4,3) & 0.8520$\pm$0.0319 & \textbf{0.8910$\pm$0.0386} & 0.8689$\pm$0.0128 & \textbf{0.8887$\pm$0.0328} & 0.8715$\pm$0.0132 \\
    \hline
  \end{tabular}
\end{table*}

\section{Effects of Segment Lengths on Seizure/Nonseizure Classification}
\label{section-effects-segment-length-on-classification}
In this section, we will explore relationship between the segment length and the performance of classifying seizures/nonseizures.

\begin{figure}[ht]
\centering
\includegraphics[width=3.2in]{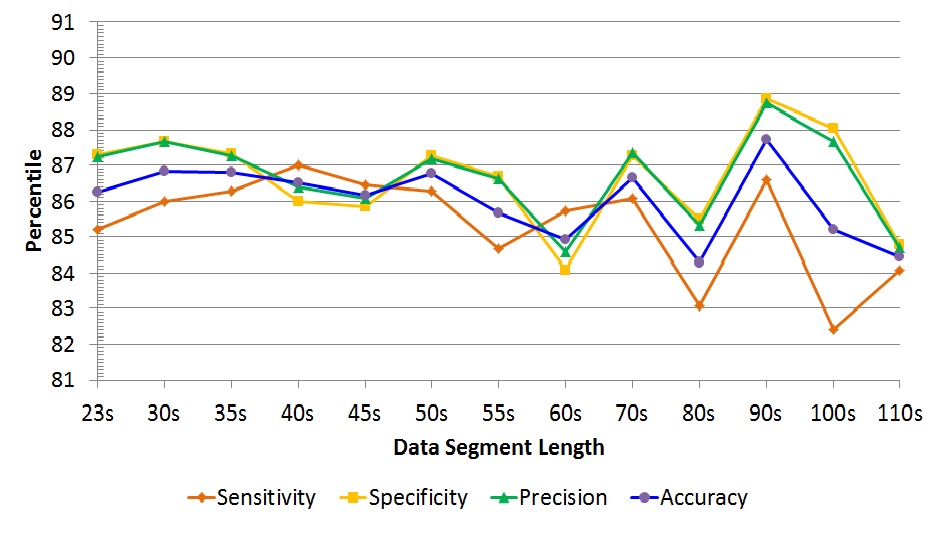}
\caption{Performance over data segments with different lengths}
\label{figure-results-over-segments-with-different-lengths}
\end{figure}

Generally, seizures last for less than two minutes. We select 13 temporal lengths less than 2 min and separately use each length to segment EEG signals in CHB-MIT. Besides the length of 23s, other 12 lengths are 30s, 35s, 40s, 45s, 50s, 55s, 60s, 70s, 80s, 90s, 100s, and 110s, respectively. Their segmentation methods are similar to the case of 23s in Section \ref{section_data_segmentation}. For each length, ten CV experiments are performed based on a group of tuned optimal parameters by using the model IndRNN-12. In the experiments for these 12 lengths, the tuned parameters mainly include the learning rate and the number of epochs, and the numbers of hidden states are set the same as in the case of 23s. The obtained CV results over the 12 different segment lengths are listed in Table \ref{expriments-over-segments-with-different-lengths}. Visualizations of these experiment results are shown in Fig. \ref{figure-results-over-segments-with-different-lengths}. The abbreviation of Seg. Len. means segment length.

\begin{table*}[htb]
\centering
\caption{CV results over segments with different lengths}
\label{expriments-over-segments-with-different-lengths}
  \begin{tabular}{|c|c|c|c|c|c|}
  \hline
    \textbf{Seg. Len.} &  \textbf{Sens.} & \textbf{Spec.} & \textbf{F1 Score} & \textbf{Prec.} & \textbf{Acc.} \\
    \hline
   23s & 0.8520$\pm$0.0424 &  0.8730$\pm$0.0377 &  0.8608$\pm$0.0175 &  0.8723$\pm$0.0314 &  0.8625$\pm$0.0157  \\
   \hline
    30s & 0.8598$\pm$0.0427 &  0.8768$\pm$0.0398 &  0.8669$\pm$0.0186 &  0.8768$\pm$0.0329 &  0.8683$\pm$0.0175  \\
    \hline
    35s & 0.8627$\pm$0.0348 &  0.8733$\pm$0.0354 &  0.8673$\pm$0.0298 &  0.8725$\pm$0.0329 &  0.8680$\pm$0.0298  \\
    \hline
    40s & \textbf{0.8700$\pm$0.0370} &  0.8600$\pm$0.0541 &  0.8659$\pm$0.0283 &  0.8640$\pm$0.0463 &  0.8650$\pm$0.0293  \\
    \hline
    45s & 0.8646$\pm$0.0236 & 0.8585$\pm$0.0429 & 0.8622$\pm$0.0220 & 0.8609$\pm$0.0374 & 0.8615$\pm$0.0236  \\
    \hline
    50s & 0.8629$\pm$0.0300 &  0.8726$\pm$0.0233 &  0.8670$\pm$0.0204 &  0.8717$\pm$0.0208 &  0.8677$\pm$0.0198  \\
    \hline
    55s & 0.8467$\pm$0.0452 &  0.8667$\pm$0.0483 &  0.8550$\pm$0.0235 &  0.8664$\pm$0.0376 &  0.8567$\pm$0.0226  \\
    \hline
    60s & 0.8574$\pm$0.0414 &  0.8407$\pm$0.0505 &  0.8503$\pm$0.0221 &  0.8457$\pm$0.0371 &  0.8491$\pm$0.0227  \\
    \hline
    70s & 0.8608$\pm$0.0310 &  0.8726$\pm$0.0450 &  0.8660$\pm$0.0135 &  0.8735$\pm$0.0365 &  0.8666$\pm$0.0153  \\
    \hline
    80s & 0.8306$\pm$0.0429 &  0.8551$\pm$0.0423 &  0.8407$\pm$0.0261 &  0.8532$\pm$0.0363 &  0.8428$\pm$0.0255  \\
    \hline
    90s & 0.8659$\pm$0.0570 &  \textbf{0.8886$\pm$0.0448} &  \textbf{0.8754$\pm$0.0392} &  \textbf{0.8874$\pm$0.0417} &  \textbf{0.8773$\pm$0.0380}  \\
    \hline
    100s & 0.8239$\pm$0.0579 &  0.8804$\pm$0.0569 &  0.8475$\pm$0.0383 &  0.8766$\pm$0.0528 &  0.8522$\pm$0.0368 \\
    \hline
    110s & 0.8409$\pm$0.0249 &  0.8477$\pm$0.0204 &  0.8437$\pm$0.0154 &  0.8470$\pm$0.0172 &  0.8443$\pm$0.0144 \\
    \hline

  \end{tabular}
\end{table*}

Fig. \ref{figure-results-over-segments-with-different-lengths} shows that the relationship between the data segment lengths and the performances of seizure/nonseizure classification is not clear-cut. The classification performance does not always go up or go down as the segment length increases, and it fluctuates over the segment lengths. With six lengths, including 60s, 70s, 80s, 90s, 100s, and 110s, the performances manifest wide fluctuating margins. With six lengths of 23s, 30s, 35s, 40s, 45s, and 50s, the differences of performances are relatively small. The best performance is obtained with the segment length of 90s. For the three metrics of Sensitivity, Specificity and Precision, their maximal gaps are all more than 4\%. For the F1 Score and Accuracy, the maximal differences are more than 3\%. It can be seen that the influence of the segment length can not be overlooked for the seizue/nonseizure classification.

Different segment lengths result in different numbers of seizure segments and different amounts of seizure data in seizure segments. The amount of seizure data can be characterized by statistics, such as the average temporal length of seizure data per seizure segment and the percentage of segments with special seizure duration in all the seizure segments.
The abbreviation of Num. Sei. Seg is for the number of seizure segments, Avg. Sei. Len. is for an average seizure duration per seizure segment, and Perc. Seg. Type-$k$ is for a percentage of Type-$k$ segments in all the seizure segments, $k=1,2,3$. A Type-1 segment is a seizure segment in which the seizure duration is less than or equal to one quarter of segment length. A Type-2 segment is a seizure segment that its seizure duration is more than or equal to one half of segment length. A Type-3 segment is a seizure segment such that the seizure duration is more than or equal to three quarters of segment length. Generally, the four statistic values, including the number of seizure segments (abbrev., Num. Sei. Seg.), the average of seizure duration per seizure segment (abbrev., Avg. Sei. Len.), the percentage of Type-2 segments (abbrev., Perc. Seg. Type-2) and the percentage of Type-3 segments (abbrev., Perc. Seg. Type-3), are the greater, the performance of seizure/nonseizure classification is the better. And the percentage of Type-1 segments (abbrev., Perc. Seg. Type-1) is the smaller, the performance is the better. Actually, as the segment length increases, seizure segments decrease and the average of seizure duration per seizure segment increases. For our selected 13 segment lengths, the statistics of seizure data are presented in Table \ref{statistic-result-of-segments-with-different-lengths}.
Besides the changes in the number of seizure segments and the average of seizure length in Table \ref{statistic-result-of-segments-with-different-lengths}, the percentage of Type-2 segments and that of Type-3 segments mostly decrease, and that of Type-1 segments increases in most cases.
According to Fig. \ref{figure-results-over-segments-with-different-lengths}, the best trade-off is achieved with the segment length of 90s for the seizure/nonseizure classification. For the segment length of 90s, the average seizure duration is the largest, and the percentage of Type-1 segments is not large relatively, around one third of seizure segments.

\begin{table*}[htb]
\centering
\caption{Statistics of segments with different lengths}
\label{statistic-result-of-segments-with-different-lengths}
  \begin{tabular}{|c|c|c|c|c|c|}
  \hline
    \textbf{Seg. Len.} & \textbf{Num. Sei. Seg.} & \textbf{Avg. Sei. Len.} & \textbf{Perc. Seg. Type-1} & \textbf{Perc. Seg. Type-2} & \textbf{Perc. Seg. Type-3} \\
    \hline
   23s & 665 & 16.99s & 14.74\%  & 76.09\% & 59.85\% \\
   \hline
   30s & 543 & 20.83s & 16.21\%  & 70.53\% & 53.41\%  \\
   \hline
   35s & 496 & 22.73s & 19.35\% & 65.93\%  & 47.18\%  \\
   \hline
   40s & 463 & 24.43s & 23.11\% & 59.83\% & 40.82\% \\
   \hline
   45s & 429 & 26.36s & 23.54\% & 59.44\%  & 35.90\% \\
   \hline
   50s & 411 & 27.52s & 25.79\%  & 49.39\%  & 32.85\% \\
   \hline
   55s & 396 & 28.49s & 28.03\%  & 48.23\%  &  29.29\%  \\
   \hline
   60s & 358 & 31.59s & 27.93\%  & 49.44\%  & 30.45\%   \\
   \hline
   70s & 340 & 33.15s & 32.65\%  & 40.29\%  & 24.41\%   \\
   \hline
   80s & 325 & 34.8s & 40.00\%  & 37.54\% & 19.69\%   \\
   \hline
   90s & 289 & 39.13s & 33.56\%  & 39.10\%  &  17.65\%   \\
   \hline
   100s & 305 & 37.08s & 47.21\% &  29.51\%  &  12.13\%  \\
   \hline
   110s & 293 & 38.32s & 48.12\%  & 25.26\%  & 10.58\%  \\
   \hline

  \end{tabular}
\end{table*}

\section{Discussion}
\label{section-discussion}
Keeping in mind a labor-intensive clinical practice that neurologists manually review off-line EEG data records for diagnosing epilepsy, we have developed an automatic approach of ADIndRNN to identify seizure segments and nonseizure segments. Each EEG data record is segmented into segments, and then the obtained segments are classified into two categories, seizure versus nonseizure. The classified seizure segments are provided to neurologists to make further analyses. For the task of classifying seizures/nonseizures to aid neurologists with off-line annotations, the metrics such as sensitivity, specificity and precision are more relevant than the metrics of false alarm rate and latency. So, we evaluate our proposed approach with five metrics, including sensitivity, specificity, F1 score, precision, and accuracy.

By integrating IndRNN with a dense structure and an attention mechanism, we propose the deep learning approach of ADIndRNN for the seizure/nonseizure classification. Our proposed approach outperforms the LSTM approach and the CNN approach with the improvements of at least 4\%. IndRNN is a variant of RNN. It supports stacking multiple layers and can handle longer sequences than LSTM and RNN\cite{Shuai_Li}. The IndRNN layer extracts features from the forward direction. The dense structure is to ensure maximum information flow between layers\cite{Gao-Huang}, and the attention mechanism is used to extract spatial features. The validation results in Section \ref{section-validation-of-modules} demonstrate that the dense structure and the attention mechanism have positive effects on the performance of our approach. In the development of the approach, a bidirectional structure of IndRNN layers is constructed in order to extract features in two directions. Our experiments show that the bidirectional structure of IndRNN is time-consuming in computation while the performance gain is marginal.

We utilize the attention mechanism to generate attention weights over channels instead of adopting direct training method. For the directly training way, parameters representing weights on channels are learned. After training, the learned weights are the same for all the data segments. In fact, one patient possibly experiences different seizure types. Seizures with different types may originate in different brain regions. And different patients may have different seizure patterns. The weights on channels, which describe the strengths of signifying seizures, need to change with seizure types and seizure patterns. So, without dwelling on the directly training method we choose to use the attention mechanism. In the attention mechanism, a kernel matrix and a bias matrix are obtained by training. The two trained matrices are combined with data segments to adaptively generate weights through transformations.

When segmenting EEG records, we attempt to ensure that the obtained data segments are close to a real-world scenario. A seizure segment could contain seizure data and nonseizure data. It is unrealistic in the real world that all the seizure segments only contain seizure data. As the LSTM approach and the CNN approach were evaluated over signals with duration of 23.6 seconds from Bonn EEG data set\cite{Acharya,Hussein}, a segment length of 23s was selected for evaluating the proposed approach against the LSTM and CNN approaches.

In the exploration of relations between segment lengths and performances of classifying seizure/nonseizure, we adopt the model IndRNN-12. The choice is based on the following three considerations: (1) For models containing more IndRNN layers or attention layer or dense structure, more parameters need to be trained, while the number of seizure segments decreases as the segment length increases, which increases the over-fitting risk. (2) The model IndRNN-12 has relatively good results over the 23s segments, as shown in Table \ref{experiment-results-for-deeper-IndRNN-related-models}. (3) The used model in the exploration needs to be kept the same for the 13 segment lengths.

\section{Conclusions}
\label{section-conclusion}
This paper considers automatical classification of seizures/nonseizures for assisting neurologists in making epilepsy diagnosis. We propose a new approach of dense IndRNN with attention by integrating an emerging neural network model, IndRNN, with a densely connected structure and an attention mechanism. The IndRNN supports stacking multiple layers to capture seizure patterns, the dense structure ensures maximum information flow between layers, and the attention mechanism helps extract spatial features. Evaluations of our proposed approach are performed in CV experiments over the noisy data set of CHB-MIT. The obtained average sensitivity, specificity, F1 score, precision and accuracy are 88.80\%, 88.60\%, 88.71\%, 88.69\%, and 88.70\%, respectively. These results exceed the LSTM approach\cite{Hussein} and the CNN approach\cite{Acharya} with an improvement of at least 4\%. Additionally, we explore how the segment length affects the performance of seizure/nonseizure classification. Our CV experiments over 13 segment lengths indicate that the classification performance fluctuates over the segment lengths, with the maximal fluctuating margin being more than 4\%. The segment length is thus an important factor influencing the seizure/nonseizure classification performance. As a future research line, we will further investigate how to use the dense IndRNN with attention for real-time seizure detection.


\begin{thebibliography}{1}

\bibitem{Fisher}
R. S. Fisher, W. V. E. Boas, W. Blume, C. Elger, P. Genton, P. Lee, et al., \lq\lq Epileptic seizures and epilepsy: Definitions proposed by the international league against epilepsy (ILAE) and the international bureau for epilepsy (IBE),\rq\rq \ \textit{Epilepsia}, vol. 46, no. 4, pp. 470-472, 2005.

\bibitem{Megiddo}
I. Megiddo, A. Colson, D. Chisholm, T. Dua, A. Nandi, R. Laxminarayan, \lq\lq Health and economic benefits of public financing of epilepsy treatment in India: An agent-based simulation model,\rq\rq \ \textit{Epilepsia}, vol. 57, no. 3, pp. 464-474, Jan, 2016.

\bibitem{Elger}
C. E. Elger, C. Hoppe, \lq\lq Diagnostic challenges in epilepsy: Seizure under-reporting and seizure detection,\rq\rq \ \textit{Lancet Neurol}, vol. 17, no. 3, pp. 279-288, Mar. 2018.

%\bibitem{Gotman1979}
%Gotman J, Ives JR, Gloor P. Automatic recognition of inter-ictal epileptic activity in prolonged EEG recordings. Electroencephalography and Clinical Neurophysiology. 1979 May;46(5):510-520.

\bibitem{Gotman1982}
J. Gotman, \lq\lq Automatic recognition of epileptic seizures in the EEG,\rq\rq \ \textit{Electroencephalography and Clinical Neurophysiology}, vol. 54, no. 5, pp. 530-540, Nov. 1982.

\bibitem{Thodoroff}
P. Thodoroff, J. Pineau, A. Lim, \lq\lq Learning robust features using deep learning for automatic seizure detection,\rq\rq \ \textit{Journal of Machine Learning Research}, vol. 56, pp. 178-190, 2016.


\bibitem{Furbass}
F. F\"{u}rbass, P. Ossenblok, M. Hartmann, et al., \lq\lq Prospective multi-center study of an automatic online seizure detection system for epilepsy monitoring units,\rq\rq \ \textit{Clinical Neurophysiology}, vol. 126, no. 6, pp. 1124-1131, 2015.

\bibitem{Zandi}
A. S. Zandi, M. Javidan, G. A. Dumont, R. Tafreshi, \lq\lq Automated real-time epileptic seizure detection in scalp EEG recordings using an algorithm based on wavelet packet transform,\rq\rq \ \textit{IEEE Trans. Biomed. Eng.}, vol. 57, no. 7, pp. 1639-1651, Jun. 2010.

\bibitem{Shoeb2010}
A. Shoeb, J. Guttag, \lq\lq Application of machine learning to epileptic seizure detection,\rq\rq \ in \textit{Proc. ICML2010}, Haifa, Israel, 2010, pp. 975-982.

\bibitem{Shoeb2009}
A. Shoeb, \lq\lq Application of machine learning to epileptic seizure onset detection and treatment,\rq\rq \ Ph. D. dissertation, Harward-MIT Division of Health Sciences and Technology,  Massachusetts Institute of Technology, Cambridge, MA, USA, 2009.

\bibitem{Kiranyaz}
S. Kiranyaz, T. Ince, M. Zabihi, D. Ince, \lq\lq Automated patient-specific classification of long-term Electoencephalography,\rq\rq \ \textit{Journal of Biomedical Informatics}, vol. 49, pp. 16-31, Feb. 2014.

\bibitem{Amin}
S. Amin, A. M. Kamboh, \lq\lq A robust approach towards epileptic seizure detection,\rq\rq \ in \textit{Proc. MLSP2016}, Salerno, Italy, 2016, pp. 1-6.

\bibitem{Hunyadi}
B. Hunyadi, M. Signoretto, W. V. Paesschen, J. A. K. Suykens, S. V. Huffel, M. D. Vos, \lq\lq Incorporating structural information from the multichannel EEG improves patient-specific seizure detection,\rq\rq \ \textit{Clinical Neurophysiology}, vol. 123, no. 12, pp. 2352-2361, Dec. 2012.

\bibitem{Esbroeck}
A. V. Esbroeck, L. Smith, Z. Syed, S. Singh, Z. Karam, \lq\lq Multitask seizure detection: Addressing intra-patient variation in seizure morphologies,\rq\rq \ \textit{Machine Learning}, vol. 102, no. 3, pp. 309-321, Mar. 2016.

\bibitem{Vidyaratne}
L. Vidyaratne, A. Glandon, M. Alam, K. M. Iftekharuddin, \lq\lq Deep recurrent neural network for seizure detection,\rq\rq \ in \textit{Proc. IJCNN2016}, Vancouver, BC, Canada, 2016, pp. 1202-1207.

\bibitem{Truong2018}
N. D. Truong, A. Nguyen, L. Kuhlmann, et al., \lq\lq Integer convolutional neural network for
seizure detection,\rq\rq \ \textit{IEEE Journal on Emerging and Selected Topics in Circuits and Systems}, vol. 8, no. 4, pp. 849-857, Dec. 2018.

\bibitem{Golmohammadi}
M. Golmohammadi, S. Ziyabari, V. Shah, et al., \lq\lq Gated recurrent networks for seizure detection,\rq\rq \ in \textit{Proc. SPMB17}, Philadelphia, PA, USA, 2017, pp. 1-5.

\bibitem{Acharya}
U. R. Acharya, S. L. Oh, Y. Hagiwara, J. H. Tan, H. Adeli, \lq\lq Deep convolutional neural network for the automated detection and diagnosis of seizure using EEG signals,\rq\rq \ \textit{Computers in Biology and Medicine}, vol. 100, no. 1, pp. 270-278, Sep. 2018.

\bibitem{Andrzejak}
R. G. Andrzejak, K. Lehnertz, F. Mormann, C. Rieke, P. David, C. E. Elger, \lq\lq Indications of nonlinear deterministic and finite-dimensional structures in time series of brain electrical activity: Dependence on recording region and brain state,\rq\rq \ \textit{Physical Review E}, vol. 64, 061907, Nov. 2001.

\bibitem{Shuai_Li}
S. Li, W. Li, C. Cook, C. Zhu, Y. Gao, \lq\lq Independent recurrent neural network: Building a longer and deeper RNN,\rq\rq \ in \textit{Proc. CVPR2018}, Salt Lake City, UT, USA, 2018, pp. 5457-5466.

\bibitem{Gao-Huang}
G. Huang, Z. Liu, L. Maaten, K. Q. Weinberger, \lq\lq Densely connected convolutional networks,\rq\rq \ in \textit{Proc. CVPR2017}, Honolulu, HI, USA, 2017, pp.2261-2269.

\bibitem{Hussein}
R. Hussein, H. Palangi, R. Ward, Z. J. Wang, \lq\lq Epileptic seizure detection: A deep learning approach,\rq\rq \ 2018, arXiv:1803.09848.

\bibitem{Fergus}
P. Fergus, A. Hussain, D. Hignett, D. A. Jumeily, K. A. Aziz, H. Hamdan, \lq\lq A machine learning system for automated whole-brain seizure
detection,\rq\rq \ \textit{Applied Computing and Informatics}, vol. 12, no. 1, pp. 70-89, Jan. 2016.

\bibitem{Nicalaou}
N. Nicalaou, J. Georgiou, \lq\lq Detection of epileptic electroencephalogram based on permutation entropy and support vector machines,\rq\rq \
\textit{Expert Systems with Applications}, vol. 39, no. 1, pp. 202-209, Jan. 2012.

\bibitem{Kharbouch}
A. Kharbouch, A. Shoeb, J. Guttag, S. S. Cash, \lq\lq An algorithm for seizure onset detection using intracranial EEG,\rq\rq \ \textit{Epilepsy \& Behavior}, vol. 22, no. 1, pp. S29-S35, Dec. 2011.

\bibitem{Bolagh}
S. N. G. Bolagh, G. D. Clifford, \lq\lq Subject selection on a Riemannian manifold for unsupervised cross-subject seizure detection,\rq\rq \ 2017, arXiv:1712.00465.

\bibitem{Ansari}
A. H. Ansari, P. J. Cherian, A. Caicedo, G. Naulaers, M. Vos, S. V. Huffel, \lq\lq Neonatal seizure detection using deep convolutional neural networks,\rq\rq \  \textit{International Journal of Neural Systems}, vol. 28, no. 0, 1850011, 2018.

\bibitem{Acharya2012}
U. R. Acharya, F. Molinari, S. V. Sree, S. Chattopadhyay, K.-H. Ng, J. S. Suri, \lq\lq Automated diagnosis of epileptic EEG using entropies,\rq\rq \ \textit{Biomedical Signal Processing and Control}, vol. 7, no. 4, pp. 401-408, 2012.

\bibitem{Zhou}
W. Zhou, Y. Liu, Q. Yuan, X. Li, \lq\lq Epileptic seizure detection using lacunarity and Bayesian linear discriminant analysis in intracranial EEG,\rq\rq \ \textit{IEEE Trans. Biomed. Eng.}, vol. 60, no. 12, pp. 3375-3381, 2013.

\bibitem{Fan}
M. Fan, C. Chou, \lq\lq Detecting abnormal pattern of epileptic seizures via temporal synchronization of EEG signals,\rq\rq \ \textit{IEEE Trans. Biomed.  Eng.}, vol. 66, no. 3, pp. 601-608, 2019.

\bibitem{Truong2017}
N. D. Truong, L. Kuhlmann, M. R. Bonyadi, J. Yang, \lq\lq Supervised learning in automatic channel selection for epileptic seizure detection,\rq\rq \ \textit{Expert Systems with Applications}, vol. 86, pp. 199-207, 2017.

\bibitem{Sergey_Ioffe}
S. Ioffe, C. Szegedy, \lq\lq Batch normalization: Accelerating deep network training by reducing internal covariate shift,\rq\rq \ in \textit{Proc. ICML2015}, Lille, France, 2015, pp. 448-456.



%\bibitem{Qu}
%Qu H, Gotman J. Improvement in seizure detection performance by automatic adaptation to the EEG of each patient. Electroencephalography and Clinical Neurophysiology. 1993 February;86(2):79-87.

%\bibitem{PhysioNet}
%Goldberger AL, Amaral LAN, Glass L, et al. PhysioBank, PhysioToolkit, and PhysioNet: components of a new research resource for complex physiologic signals. Circulation. 2000 Jun 13;101:e215-e220.


\end{thebibliography}
\end{document}